 \documentclass[preprint2]{aastex}

\shorttitle{A Brown Dwarf Companion to V471 Tau}
\shortauthors{Guinan et al.}

\begin{document}

\title{The Best Brown Dwarf Yet?: A Companion to the Hyades Eclipsing Binary 
V471 Tau}

\author{Edward F. Guinan and Ignasi Ribas}
\affil{Department of Astronomy \& Astrophysics, Villanova University, 800
       Lancaster Av., Villanova, PA 19085, USA}

\begin{abstract}
We have carried out an analysis of about 160 eclipse timings spanning over 30
years of the Hyades eclipsing binary V471 Tauri that shows a long-term
quasi-sinusoidal modulation of its observed eclipse arrival times. The O--Cs
have been analyzed for the ``light-time'' effect that arises from the
gravitational influence of a tertiary companion. The presence of a third body
causes the relative distance of the eclipsing pair to the Earth to change as it
orbits the barycenter of the triple system. The result of the analysis of the
eclipse times yields a light-time semi-amplitude of $137.2\pm12.0$~s, an
orbital period of $P_3=30.5\pm1.6$~yr and an eccentricity of $e_3=0.31\pm0.04$.
The mass of the tertiary component is $M_3\sin
i_3\simeq0.0393\pm0.0038$~M$_{\odot}$ when a total mass of
$1.61\pm0.06$~M$_{\odot}$ for V471 Tau is adopted. For orbital inclinations
$i_3\ga35$\arcdeg, the mass of the third body would be below the stable
hydrogen burning limit of $M\approx0.07$ M$_{\odot}$ and it thus would be a
brown dwarf. In the next several years (near maximum elongation), it should be
feasible to obtain IR images and spectra of V471~Tau~C that, when combined with
the known mass, age, distance, and $[Fe/H]$, will serve as a benchmark for
understanding the physical properties and evolution of brown dwarfs.
\end{abstract}


\section{Introduction}

Over a hundred brown dwarf (BD) candidates have now been discovered primarily
from IR surveys such as 2MASS, DENIS, SDSS, and others (see Kirkpatrick et al.
2000). These objects have been classified as BDs from their spectra (e.g.,
presence of lithium and methane) and IR colors. A spectroscopic classification
scheme recently proposed by Kirkpatrick et al. (1999) divides these objects
into T-type (cooler BDs, methane BDs) and L-type (hotter BDs). A number of
different models have been published to explain the photometric and
spectroscopic characteristics of BDs (e.g., Burrows et al. 1997; Chabrier et
al. 2000) but a unified scenario is yet to be provided. Unfortunately, nearly
all of the BD candidates are free-floating objects in which it is not possible
to determine dynamical masses. Therefore, the major problem confronting these
models is the paucity of information on the physical properties of the BDs.
For example, the ages and luminosities of solitary BDs in nearby clusters
(e.g., Pleiades) are independently known, as well as their spectroscopic and
photometric properties, but, regrettably, no direct measures of the masses of
the candidates (the primary variable that determines the properties and the
evolution of any object) are possible. 

For a dynamical mass of a BD to be known, it has to be a member of a binary or
multiple system. There are only a handful of cases where BDs have been imaged
around nearby stars (see Burgasser et al. 2000), the most well-known of these
is Gliese 229B. This wide ($\sim$8\arcsec) astrometric binary has the first
{\em bona fide} cool (T-type) BD (Nakajima et al. 1995), and has been studied
by many observers. However, the long orbital period of Gl 229B would require
centuries to determine dynamically (i.e., directly) a definitive mass. Even
though the precision of current Doppler techniques in planet search surveys is
higher than needed for the discovery of BDs, only a few candidates have been
found so far.  This indicates a probable low frequency of short period BDs
orbiting main sequence stars (Halbwachs et al. 2000). In any case, these
objects fail to meet the requirements for being critical to understanding the
physics and structure of BDs because: their observed masses are based on
model-dependent calibrations, the orbital inclination is not known, and, also,
the glare of the bright component makes direct imaging impossible.

In addition to direct imaging and radial velocity techniques, eclipsing
binaries showing a ``light-time'' effect offer an opportunity for detecting
low-mass companions. The light-time effect occurs as the relative distance (and
light travel time) changes as the eclipsing binary moves around the barycenter
of the triple system. The detection of low-mass objects orbiting eclipsing
binary stars is of great interest because the mass of the tertiary component
can be directly determined (the analysis of the light and radial velocity
curves of the eclipsing binary yields fundamental masses for the stars). Also,
it is often possible to estimate the age of the system via isochronal fits or
using other indicators such as cluster membership. If direct imaging of the
tertiary component was possible (long period orbit), the orbital inclination,
the mass, the age, the photometry, and the spectroscopy of the third component
could all be simultaneously obtained. 

The 9th mag Hyades eclipsing binary V471 Tauri (BD+16$^{\circ}$516; HIP17962;
$P=12.5$~hr) has been under intense study at nearly all wavelengths since its
discovery in 1969 (Nelson \& Young 1970). V471 Tau is a detached binary that
consists of a hot DA2 white dwarf and a chromospherically active K2~V star
(Guinan \& Sion 1984). This system plays a key role in deciphering binary star
evolution (post common envelope and pre-cataclysmic variable phases) as well
providing fundamental information on the formation, structure and evolution of
white dwarfs. V471 Tau also serves as a laboratory for studying dynamos and
magnetic activity of cool stars. For example, during the primary eclipse, the
hot white dwarf acts as a beaming probe of the K2 star's atmosphere (see Guinan
et al. 1986). In this {\it Letter}, we present yet another aspect that
demonstrates the importance of V471~Tau. From the analysis of over 30 yr of
eclipse timings, V471 Tau shows periodic variations of the arrival times of its
primary eclipse.  Such variations appear to arise from the light-time effect
caused by the gravitational pull of a low mass tertiary companion that could
likely be a BD with a mass of $\sim0.04$~M$_{\odot}$.

\section{Detection of Third Bodies in Eclipsing Binary Systems}

A significant number of eclipsing binaries have been found to have nearby,
unseen tertiary components using light-time effects (see Demircan 2000). From
this technique, the eclipses act as an accurate clock for detecting subtle
variations in the distance to the object (this is analogous to the method used
for discovering earth-sized objects around pulsars, see Wolszczan \& Frail
1992). The periodic quasi-sinusoidal variations in the eclipse arrival times
have a very simple and direct physical meaning: the total path that the light
has to travel varies periodically as the eclipsing pair moves around the
barycenter of the triple system. The amplitude of the variation is proportional
both to the mass and to the period of the third body, as well as to the orbital
inclination. Because of selection effects, most of the detected tertiary
companions around eclipsing systems are stellar-mass objects. However,
uncomplicated short-period eclipsing binaries with sharp, well-defined
eclipses, for which very accurate arrival times can be determined, offer the
possibility of discovering low-mass tertiary components.

In the case of V471 Tau, the times of the mid-primary eclipse have been
measured with uncertainties of typically 5--10 seconds, making it possible to
discern orbital variations of the eclipsing pair with a semi-major axis as
small as 0.02 AU ($\sim$10 light-seconds). Several period studies of V471 Tau
have been made. A variation in the apparent period was pointed out by Beavers,
Lui, \& Herczeg (1986), who suggested that the deviation in the O--Cs
(observed--computed) from a linear ephemeris might be due to a three-body orbit
with a period of about 25 years. This study was superseded by the analysis of
many more eclipse timings up to 1993/4 carried out by \.Ibano\^glu et al.
(1994). With this more extensive data set, these authors estimated a period of
about $22$ years and a minimum mass of $0.038$~M$_{\odot}$ for the third body.
In the present study (preliminary results presented by Guinan et al. 2000), the
number of eclipse timings has yet been increased and the analysis has been 
refined, but, more importantly, the time-span is nearly the orbital period and
the evidence for a low mass third body is now very strong.

\section{Observations}

\begin{figure}
\epsscale{1.00}
\plotone{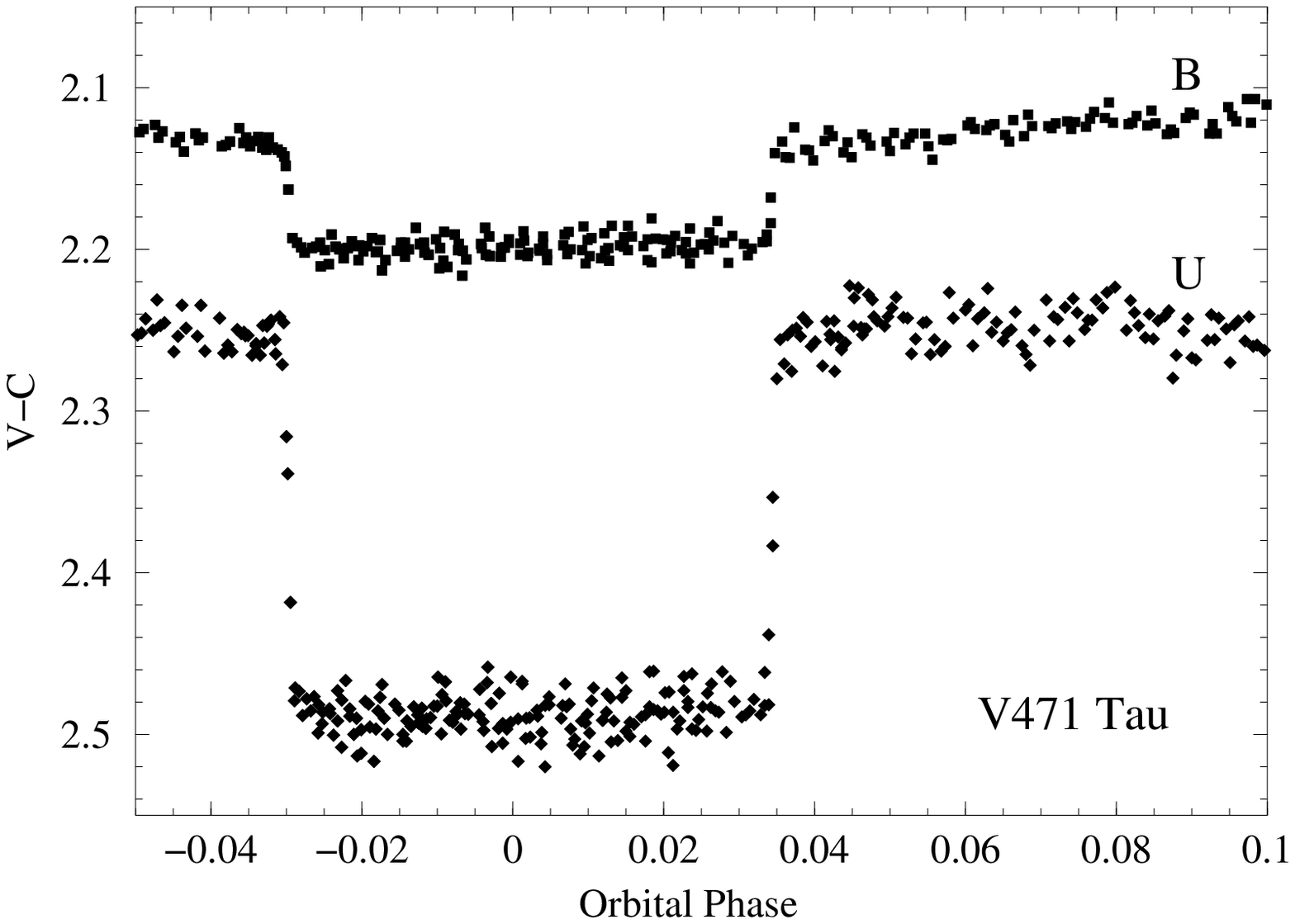}
\caption{Detail of the primary eclipse of V471~Tau. Note the
strong wavelength dependence caused by the large temperature difference
between the components. \label{fig1}}
\end{figure}

Photoelectric photometry of V471 Tau has been carried out almost continuously
since its discovery as an eclipsing binary in 1969. The light curves show some
variability due to the changing aspects and areal coverage of star-spots on the
cool star (see \.Ibano\^glu et al. 1994). The primary eclipse can easily be seen
in the $U$ and $B$ light curves from its characteristic sharp and short ingress
and egress phases (1st--2nd and 3rd--4th contacts) that last only 55~s as the
small white dwarf passes through the partial phases of its eclipse when
disappearing or appearing behind the K2 star. From our observations we measure
the total duration of the primary eclipse (i.e., 1st--4th contacts) to be
49.4~min, nearly identical to the values published since the discovery
(\.Ibano\^glu 1978). Typical primary eclipse photometry of V471~Tau is shown in
Figure \ref{fig1}. The secondary eclipse (when the white dwarf transits the
larger, cooler star) is undetectable, but the radial velocity curve indicates a
circular orbit. 

Our eclipse timing observations extend the list provided by \.Ibano\^glu et al.
(1994) up to late 2000. The observations were obtain with the Four College
Consortium 0.8-m Automatic Photoelectric Telescope located in Southern Arizona.
Differential photometry was carried out using $UBV$ or Str\"omgren $uvby$
filter sets. The observations during 1994/95, 1995/96, and early 1999 were
primarily devoted to studying the overall behavior of the light curve and not
aimed at securing individual eclipse timings. However, by phasing the data,
excellent epochal mid-primary eclipse times could be determined. During late
1999 and 2000, longer observing runs were used to obtain photometry covering
individual primary eclipses. The mid-times of primary minimum and the O--Cs for
these are given in Table \ref{tab1}, along with the corresponding
uncertainties. The O--Cs were computed using a refined ephemeris determined 
from our analysis (see \S \ref{analys}, Eq. \ref{ephem}). 

Our observations were combined with those compiled by \.Ibano\^glu et al.
(1994) to yield a total of 163 eclipse timings from 1970 through October 2000.
Such a long time span (30 yr) and the small scale of the observed effect (O--Cs
of the order of seconds) raise some concern on the uniformity of the time-scale
used for the mid-eclipse events. Even though it is not explicitly mentioned in
any of the publications, the times listed are commonly assumed to be in the UTC
(coordinated universal time) scale. As it is well known, however, UTC is not
uniform as it is affected by the fluctuations of Earth's rotation.  In the last
30 yr, the difference between UTC and a uniform time scale has been steadily
increasing at the rate of roughly 1 s/yr. Because of this, we decided to
transform all UTC timings to the TDB (barycentric dynamical time) scale using
the tables and expressions given in The Astronomical Almanac (Larsen \&
Holdaway 2000). It should be noted that the difference TDB--UTC increased from
about +40 s in 1970 to +64 s in 2000. The timings listed in Table \ref{tab1}
are therefore HJD but in the TDB scale.

\section{Analysis} \label{analys}

The expressions that describe the light-time effect as a function of the
orbital properties were first provided by Irwin (1959). In short, the time
delay or advance caused by the influence of a tertiary component can be
expressed as:
\begin{equation} \label{eq1}
\Delta T = \frac{a_{12} \sin i_3}{c} \left[\frac{1-e_3^2}{1+e_3 \cos \nu_3}
\sin(\nu_3 + \omega) + e_3 \sin \omega \right]
\end{equation}
where $a_{12}$, is the semi-major orbit of the eclipsing pair around the 
barycenter, $i_3$, and $e_3$ are the inclination, and the eccentricity of the 
third body orbit, respectively, $c$ is the speed of light, $\omega$ is
the argument of the periastron, and $\nu_3$ is the true anomaly of the third
body as it moves around the barycenter (i.e., function of time). As is
customary, the orbital inclination $i_3$ is measured relative to the plane of
the sky. Equation (\ref{eq1}) was slightly modified in order to fit the O--C
data. Indeed, we considered the projected orbital semi-amplitude in seconds
(i.e., $a_{12}\,c^{-1}\, \sin i_3$), and the true anomaly, function of the
period and the time of periastron passage, was computed to high accuracy 
using an iterative scheme. In addition to this, a zero point (to refer the
ephemeris to the barycenter of V471 Tau) and a correction to the orbital period
of the eclipsing pair (that could lead to a linear increase or decrease of the
O--Cs) were also considered. The initial values of the period and the zero
point were adopted from Bois, Mochnacki, \& Lanning (1988). An algorithm
especially designed for nonlinear functions was built to search for the orbital
properties that yield the best fit to the observations.

\begin{figure*}[!b]
\epsscale{2.30}
\plottwo{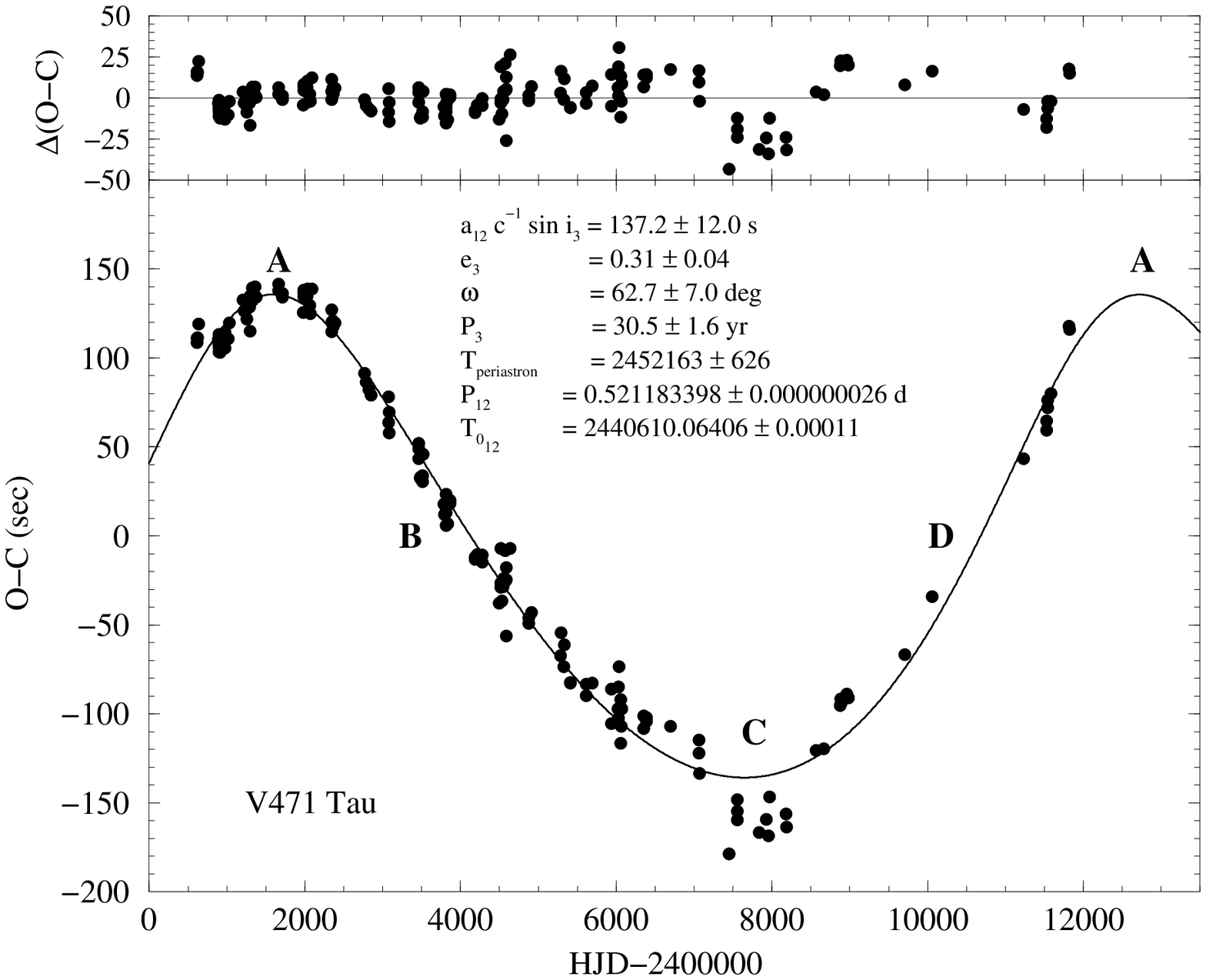}{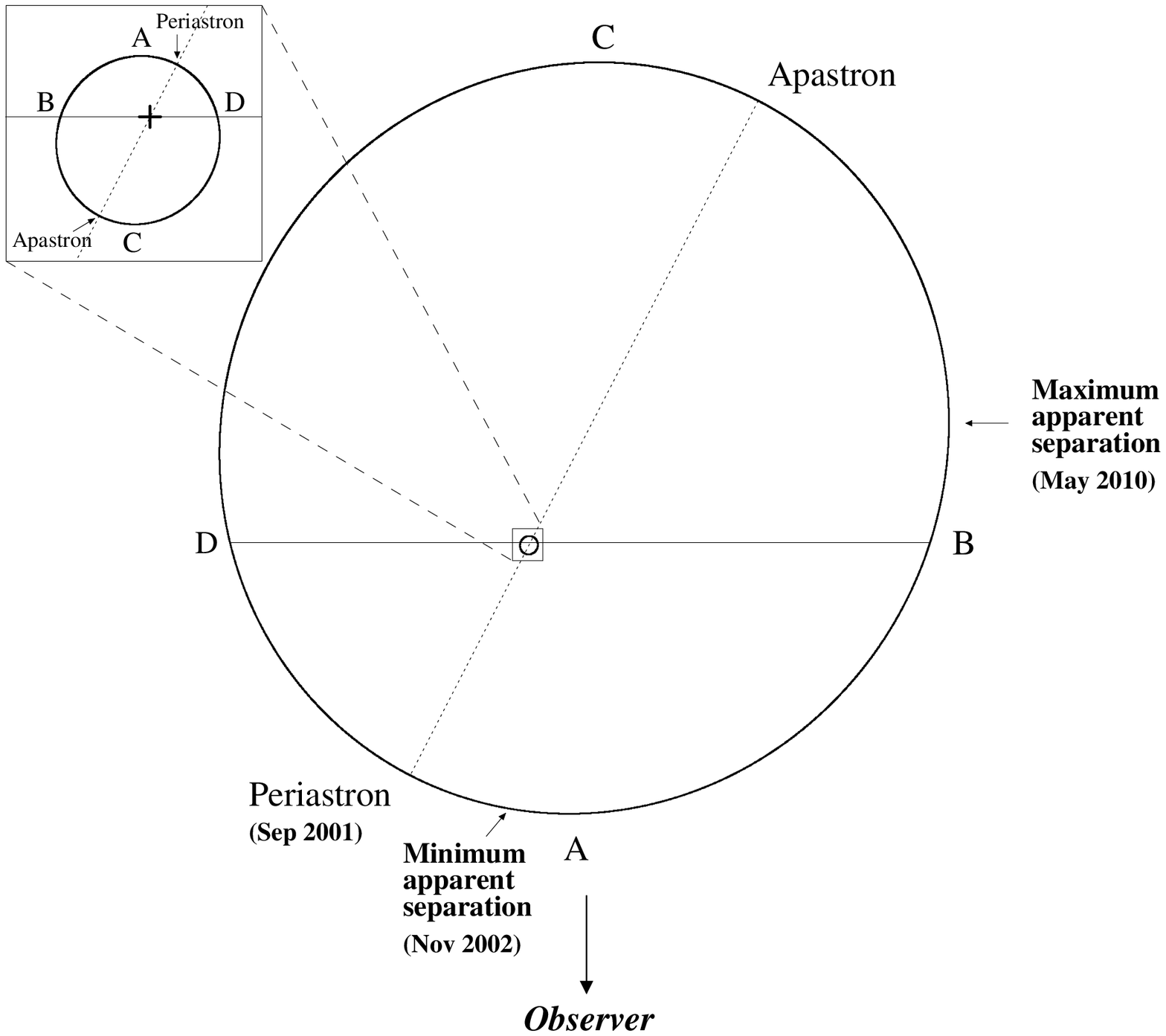}
\caption{{\em Lower left}: Plot of the residuals of the eclipse timings
according to the linear ephemeris of Eq. (\ref{ephem}). The best-fitting curve
assuming a third body perturbation is shown as a solid line.  {\em Upper left}:
Residuals after subtracting the third body perturbation. {\em Right}:
Perpendicular view to the plane of the orbit of the tertiary component.  The
letters mark specific phases also shown in the O--C plot. Calendar dates for
some important orbital phases are also provided. \label{fig2}}
\end{figure*}

Excellent fits to the eclipse arrival times were obtained, with residuals in
agreement from that expected from the timing uncertainties. However, cyclic
(not periodic) O--C variations that can mimic light-time effects are known for
some mass-exchanging semi-detached and contact binaries (see, e.g., Kreiner \&
Ziolkowski 1978). V471 Tau is a detached system with no evidence of significant
mass exchange or mass loss. Furthermore, the excellent light-time theoretical
fits to the O--C variations would be extremely unlikely if these were caused by
other mechanisms.  The best-fitting parameters are presented in Table
\ref{tab2}.  More illustrative is Figure \ref{fig2} (left), that shows a plot
of the linear ephemeris O--Cs with the best-fitting light-time curve
superimposed, and also an upper panel with the residuals of the fit. 

The modeling yields an orbital period for the third body of $P_3=30.5\pm1.6$~yr
and a light-time semi-amplitude of $a_{12}\,c^{-1}\,\sin i_3=137.2\pm16.0$~s.
The mass function of the system is therefore $f(m)=(2.23\pm0.63)\;10^{-5}$
M$_{\odot}$, that implies a minimum mass of $0.039$~M$_{\odot}$ for the third
body. The orbital eccentricity was found to be $e_3=0.31\pm0.04$ and the value
of the argument of periastron $\omega=62\fdg7\pm7\fdg0$. New accurate
ephemeris for the eclipsing pair were also determined:
\begin{equation} \label{ephem}
\mbox{Min I} = {\rm HJD}2440610.06406 + 0.521183398 \; E
\end{equation}
It should be noted that that the period found is about 5 milli-seconds longer
than the one used by \.Ibano\^glu et al. (1994) and adopted from Bois et al.
(1988). Also, the zero epoch of Bois et al. (1988) has been corrected to the
barycenter of V471 Tau and listed in the TDB scale (not UTC). This should be
kept in mind when using Eq. (\ref{ephem}).

Except for the period of the tertiary component (very much improved in the
current analysis due to the longer time base-line), our results are in general
good agreement with those published by Beavers et al. (1986) and \.Ibano\^glu
et al. (1994). Figure \ref{fig2} (right) shows a scale model of the appearance
(perpendicular to the plane of the sky) of the triple system in the case of
$i_3=90^{\circ}$. The small inner ellipse corresponds to the orbit of the
eclipsing pair, and the outer ellipse represents the orbit followed by the
tertiary component.

In addition to the 30-year period in the O--C diagram, there is also some
evidence of shorter-period (several years), lower-amplitude ($\sim20$ s) cyclic
oscillations in the eclipse timings (see Figure \ref{fig2}, upper left panel).
It is uncertain what produces these. They could arise from small period changes
of the binary itself caused by an activity cycle (\.Ibano\^glu et al. 1994) or,
if strictly periodic, from the presence of additional low-mass bodies orbiting
the binary pair. However, we carried out a further fit by giving zero weight 
to the clumps of eclipse timings around HJD2440600 and HJD2448000 that had
clearly high residuals. The result of this test was a nearly perfect fit with
no significant oscillations in the remaining residuals. The fit yielded a
semiamplitude of $170$ s, an orbital eccentricity of $0.40$, and an orbital
period of $33.8$ yr, but all these quantities had rather high uncertainties
that made them compatible with the values listed in Table \ref{tab2}. If this
were indeed the correct solution, the minimum mass of the third body would then
be 0.045~M$_{\odot}$. Additional observations of eclipse timings (especially
critical near the maximum of the O--C curve) will reduce the current
uncertainties and firmly establish the orbital properties of the tertiary
component.

\section{Discussion}

\begin{figure}
\epsscale{1.00}
\plotone{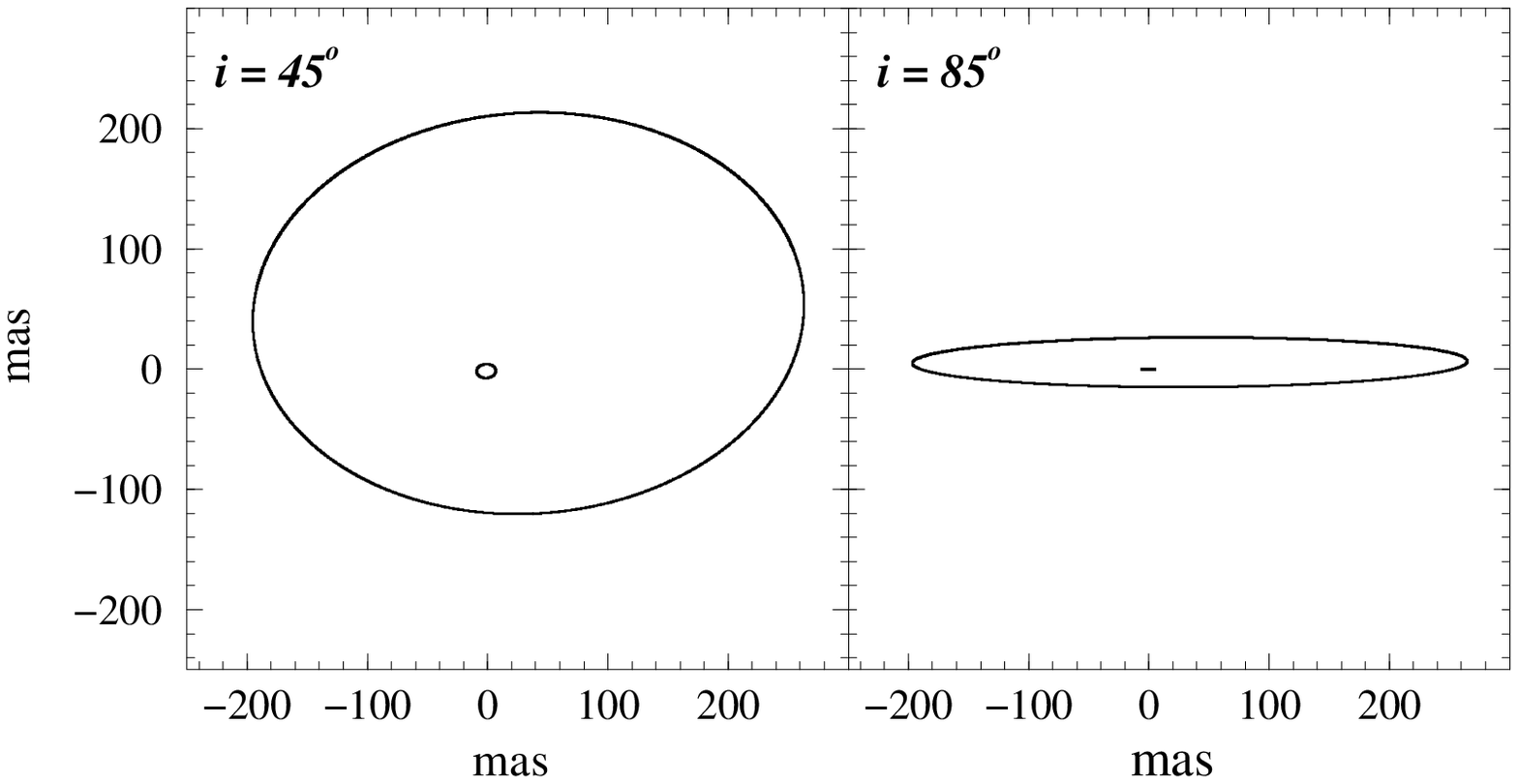}
\caption{Scale projection in the plane of the sky of the orbits 
of the eclipsing pair of V471~Tau (small inner ellipse) and the tertiary
component (larger ellipse). The angular scale has been computed by adopting
a distance of $d=46.8$~pc for the system. \label{fig3}}
\end{figure}

The results presented in the previous section yield strong evidence for the
presence of a tertiary component in a long-period orbit around V471~Tau.
Adopting a total mass for the eclipsing binary of $M_1+M_2=
1.61\pm0.06$~M$_{\odot}$ ($M_{\rm K2}=0.79\pm0.04$~M$_{\odot}$, $M_{\rm
WD}=0.82\pm0.04$~M$_{\odot}$; E. M. Sion, priv. comm.), the mass of the third
body is $M_3 \sin i_3\simeq0.0393\pm0.0038$~M$_{\odot}$ and its orbital
semi-major axis is $a_3\simeq11.2\pm0.4$~AU. The mass of the third component as
a function of the inclination (for several selected values) is provided in
Table \ref{tab3}. As can be seen, as long as the inclination of the orbital
plane of the tertiary component is $\ga35$\arcdeg, the mass of this object is
below the theoretical threshold of $\sim0.07$~M$_{\odot}$ for a hydrogen
burning star. Even for the minimum mass, however, the third body would be
probably too massive to be a planet. The predicted orbit projected on the sky
as a function of the orbital inclination has also been calculated. Plots for
two different values of $i_3$ are shown in Figure \ref{fig3}. In these
calculations a distance of $46.8\pm3.5$~pc determined from Hipparcos astrometry
of V471 Tau (ESA 1997) has been adopted. The maximum apparent separation of the
tertiary component from the eclipsing pair reaches $\approx 260$~mas near one
of the quadratures of its orbit. 

Thus, according to our calculations, V471~Tau~C is likely to be a BD for which
the mass, age, and chemical composition can be well known. The age and chemical
composition of the system are known from its membership in the Hyades cluster
(Barstow et al. 1997). Recent determinations of the Hyades age yield a value of
$625\pm50$~Myr (Perryman et al. 1998). From its mass, age, and $[Fe/H]$, most
of the observable properties of the object can be estimated from theoretical
models. We adopted very recent evolutionary models of Chabrier et al. (2000)
that consider non-gray atmospheres with dust formation and opacity. The
predictions for several values of $i_3$ (obtained from a double linear
interpolation in age and mass) indicate values of $T_{\rm eff}=1370$~K, $\log
g=5.0$, and $M_{\rm K}=12.7$~mag for the minimum mass of the third body
(see Table \ref{tab3}). The apparent K-band magnitude of the low-mass object
would be $m_{\rm K} \la 16.0$ mag. The BD companion is therefore expected from
theory to be $\approx9$~mag fainter than the eclipsing binary in the K-band
(V471 Tau: $m_{\rm V}=9.43\pm0.05$~mag, $m_{\rm K}\simeq7.21\pm0.10$~mag).

We have proposed HST/{\em Fine Guidance Sensors} astrometry (to be combined
with existing Hipparcos data) to ascertain the presence and determine the
properties of the tertiary component. When these astrometric measurements are
available (covering half of the orbital period) they will unambiguously yield
the orbital inclination and the semi-major axis with an error below 0.5 mas,
corresponding to a few percent uncertainty of the tertiary object's mass. This
will represent the first direct dynamical mass determination of a BD with known
age, chemical composition, and distance. Moreover, in the next several years it
should be feasible to image directly V471 Tau C in the IR (using coronographic
observations made with adaptive optics or observations made from space) as it
moves toward maximum angular elongation from the eclipsing pair. Once we carry
out these observations, and if V471 Tau C is indeed confirmed to be a BD, it
will make an excellent benchmark for understanding the properties and evolution
of BDs.

\acknowledgments

The APT observations were acquired and reduced using programs developed by G.
P. McCook, who is gratefully acknowledged. Villanova student J. J. Bochanski is
thanked for his help during the early stages of this work. I. R. acknowledges
the Catalan Regional Government (CIRIT) for financial support through a
postdoctoral Fulbright fellowship. This research was supported by NSF/RUI
grants AST 93-15365, AST 95-28506, and AST 00-71260, and NASA/HST grant
GO-6088.04.

\clearpage


\begin{deluxetable}{clcr}
\tablewidth{0pt}
\tabletypesize{\footnotesize}
\tablecaption{New primary eclipse timings for V471 Tau.
\label{tab1}}
\tablehead{\colhead{HJD\tablenotemark{a}~$-$2400000} & \colhead{O--C (s)} &
\colhead{HJD\tablenotemark{a}~$-$2400000} & \colhead{O--C (s)}}
\startdata
49703.67121\tablenotemark{b}&$-$67$\pm$10 & 
51538.75956                 &  +72$\pm$6  \\
50061.72458\tablenotemark{b}&$-$34$\pm$5  & 
51538.75961                 &  +76$\pm$8  \\
51229.69748\tablenotemark{b}&  +43$\pm$13 & 
51584.62379                 &  +80$\pm$9  \\
51528.85693                 &  +59$\pm$9  & 
51812.90256                 & +118$\pm$5  \\
51528.85699                 &  +65$\pm$7  &
51823.84739                 & +116$\pm$5  \\
\enddata
\tablenotetext{a}{Not in the UTC but in the TDB scale (see text).}
\tablenotetext{b}{Seasonal average.}
\end{deluxetable}

\begin{deluxetable}{lr}
\tablewidth{0pt}
\tablecaption{Orbital solution for the tertiary component of V471 Tau.
\label{tab2}}
\tablehead{\colhead{Parameter} & \colhead{Value}}
\startdata
Period (yr)                   & $30.5\pm1.6$     \\
Semiamplitude (s)             & $137.2\pm12.0$   \\
Eccentricity                  & $0.31\pm0.04$    \\
$\omega$ (\arcdeg)            & $62.7\pm7.0$     \\
$T_{\rm periastron}$ (HJD)    & $2452163\pm626$  \\
Period EB (d)                 & $0.521183398\pm0.000000026$\\
$T_{\circ}$ EB (HJD)          & $2440610.06406\pm0.00011$\\
\enddata
\end{deluxetable}

\begin{deluxetable}{clccccc}
\tabletypesize{\footnotesize}
\tablewidth{0pt}
\tablecaption{Predicted properties of the tertiary component of V471 Tau 
as a function of the orbital inclination. \label{tab3}}
\tablehead{
\colhead{$i_3$ ($^{\circ}$)}&
\colhead{$M/$M$_{\odot}$}&
\colhead{$T_{\rm eff}$ (K)}&
\colhead{$\log (L/{\rm L}_{\odot})$}&
\colhead{$J$$-$$K$}&
\colhead{$m_{\rm V}$}&
\colhead{$m_{\rm K}$}}
\startdata
85&0.039&1370&$-$4.5&4.1&32.8&16.0\\
60&0.045&1540&$-$4.3&3.1&29.6&15.4\\
45&0.056&1830&$-$4.0&1.8&25.8&14.7\\
30&0.080&2460&$-$3.4&0.8&21.3&13.7\\
25&0.095&2730&$-$3.1&0.7&19.4&13.0\\
\enddata
\tablecomments{Based upon non-gray dusty models of Chabrier et al. 2000. An 
age of 625~Myr has been adopted for the system from its Hyades membership.  The
apparent magnitudes have been computed by adopting a distance of $d=46.8$~pc
for the system.}
\end{deluxetable}

\clearpage 

\end{document}